\documentclass[aps,prb,showpacs,groupedaddress,twocolumn]{revtex4}
\usepackage{amsmath,amssymb}
\usepackage{graphicx} 
\usepackage{dcolumn} 
\usepackage{bm} 

\begin{document}
\draft
\title{Disorder induced quantized conductance with fractional value and universal conductance fluctuation in three-dimensional topological insulators}
\author{Lei Zhang$^{1}$, Jianing Zhuang$^{1}$, Yanxia Xing$^{1,2}$ and Jian Wang$^{1,\ast}$}
\address{$^1$Department of Physics and the Center of Theoretical and
Computational Physics, The University of Hong Kong, Hong Kong,
China\\$^2$Department of Physics, Beijing Institute of Technology, Beijing
100081, China}


\begin{abstract}
We report a theoretical investigation on the conductance and its
fluctuation of three-dimensional topological insulators (3D TI) in
$Bi_2Se_3$ and $Sb_2Te_3$
in the presence of disorders.
Extensive numerical simulations are carried out. We find that in the
diffusive regime the conductance is quantized with fractional value.
Importantly, the conductance
fluctuation is also quantized with a universal value. For 3D TI
connected by two terminals, three independent conductances $G_{zz}$,
$G_{xx}$ and $G_{zx}$ are identified where z is the normal
direction of quintuple layer of 3D TI (see inset of Fig.1). The
quantized conductance are found to be $\langle G_{zz}\rangle=1$, $\langle G_{xx}\rangle=4/3$
and $\langle G_{zx}\rangle=6/5$ with corresponding quantized conductance
fluctuation $0.54$, $0.47$, and $0.50$. The quantization of average
conductance and its fluctuation can be understood by theory of mode mixing. The
experimental realization that can observe the quantization of
average conductance is discussed.
\end{abstract}
 \maketitle

Recently, the topological insulator (TI), a new state of matter, has attracted a lot of
theoretical and experimental attention.\cite{moore,kane2,zhang2} The TI has an insulating energy gap in the bulk
states which behaves like the general insulator, but it has exotic gapless metallic states
on its edges or surfaces. The TI is first predicted in two-dimensional (2D) systems, e.g.,
the graphene and HgTe/CdTe quantum well. It has been generalized\cite{kane1} in 3D and confirmed
experimentally.\cite{hsieh} The 2D TI has the gapless helical edge states and exhibits
the quantum spin Hall effect while in 3D TI the conducting state is
helical surface state. This helical edge or surface states are topologically protected
and are robust against all time-reversal-invariant impurities. Many interesting physical phenomena have been predicted including Majorana fermion\cite{majorana}, topological magnetoelectric effect\cite{zhang1}, magneto-optical Kerr and Faraday effects.\cite{tse} There are also many studies on disordered TI. It was found by Li et al that in the presence
of disorder,\cite{li} a new phase called topological Anderson insulator can be induced.
The complete physical picture and mechanism of topological Anderson insulator (TAI) was
given by Groth et al\cite{Groth} using an effective medium theory. Recently, the TAI was also
predicted in 3D TI.\cite{guo} Both analytic and numerical results show that disorder can induce strong
topological Anderson insulator in 3D.

Disorder can affect mesoscopic systems in an important way. For instance, it is well
known that, in the diffusive regime, the conductance fluctuation of the mesoscopic system
assumes a universal value that is independent of system parameters and depends only
on dimensionality and symmetry of the system.\cite{ucf} For a 2D TI (HgTe/CdTe quantum well), it was found that the spin-Hall conductance fluctuation is universal.\cite{qiao} It will be of great interest to explore the universal behavior of conductance and its fluctuation in 3D TI.

\begin{figure}
\includegraphics[width=8.5cm,height=5.5cm]{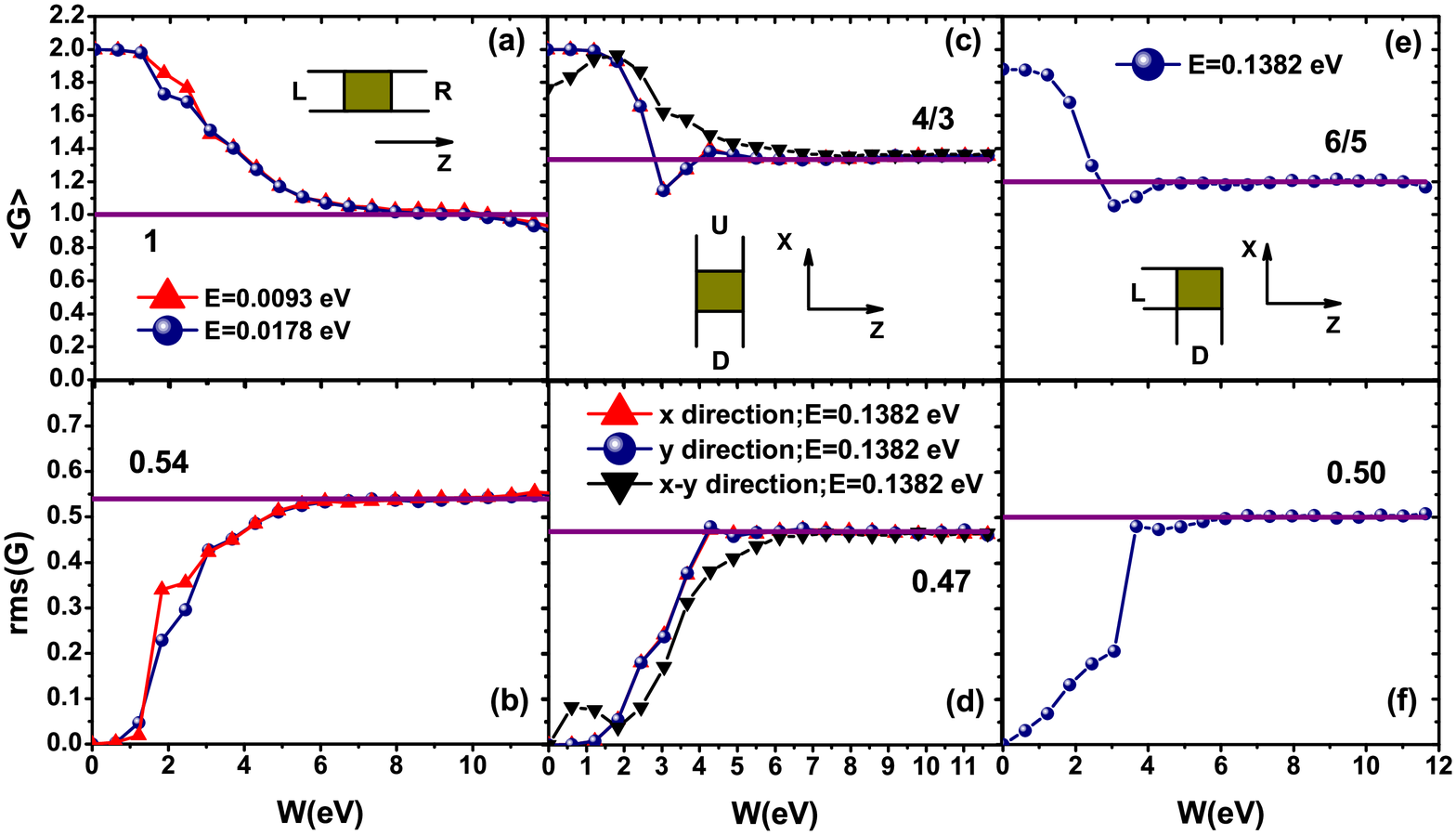} \caption{
(Color online)Conductance and its fluctuation vs disorder strength
for different transport directions, $G_{zz}$ for panel $(a)$ and
$(b)$, $G_{xx}$, $G_{yy}$ and $G_{xy}$ for panel $(c)$ and $(d)$, $G_{xz}$ for
panel $(e)$ and $(f)$. Each data point on the figure is averaged over 5000
configurations.\label{fig1}}
\end{figure}

In this paper, we have studied the effect of disorder on topological surface states in 3D TI for $Bi_2Se_3$ system using extensive numerical simulation. Due to anisotropy of the $Bi_2Se_3$ system, there are three independent two-terminal conductance $G_{zz}$, $G_{xx}$ and $G_{zx}$ where z is the normal direction of quintuple layer of 3D TI. We found that in the diffusive regime, the average conductance is quantized with fractional value $1,4/3,6/5$, respectively, for $G_{zz}$, $G_{xx}$ and $G_{zx}$ (in units of $e^2/h$). The corresponding conductance fluctuation is also quantized. By varying the disorder strength $W$ and the Fermi energy $E_F$ inside the bulk gap, the phase diagrams of conductance and its fluctuation in the plane of $(W,E_F)$ were constructed, showing the same quantized conductance and fluctuation for a wide range of $E_F$ and $W$. By introducing the number of effective transmission and reflection channels, the quantized conductance as well as the quantized fluctuation can be understood by the mode mixing theory.\cite{Levitov} The quantized values of conductance and fluctuation from the proposed effective mode mixing theory are in excellent agreement with our numerical results. It is known that as disorder increases the electron transport goes through the ballistic and the diffusive regime and finally enters the localized regime. Our numerical result shows that the evolution of topological surface states in the presence of disorders undergoes three stages: (1). the topological surface states are protected up to a critical disorder strength beyond which the topological surface states are destroyed while the bulk gap is present. At this stage, the conducting channels are non-topological surface states. This is true when the system begin entering the diffusive regime. (2). While still in the diffusive regime but at large disorders, the bulk gap is destroyed and the conducting channels are bulk states. (3). For large enough disorders, the system becomes localized and there is no conducting channel available. Furthermore, four-terminal conductance and its fluctuation were calculated. Similar quantization behaviors for conductance and fluctuation were found.

In our numerical simulation, we discretize spatial
coordinates of the continuous effective $\mathbf{k}$ space
Hamiltonian $H_{TI}$ in Ref.\onlinecite{Zhang} on the square
lattice:
\begin{eqnarray}
H_{TI} = \sum_{{\bf i}}\Psi^{\dagger}_{{\bf i}}H_{{\bf ii}}\Psi_{{\bf i}}
+\sum_{\vec{\alpha}=(\vec{\delta} x,\vec{\delta} y,\vec{\delta} z),{\bf i}}\Psi^{\dagger}_{{\bf i}}
H_{{\bf i},\vec{\alpha}}\Psi_{{\bf i}+\vec{\alpha}}+\emph{H.c.}\label{Ham}
\end{eqnarray}
where ${\bf i}=(ix,iy,iz)$ is the site index and $\vec{\delta}
x,\vec{\delta} y,\vec{\delta} z$ are unit vectors along $x,y$ and
$z$ directions where z is the normal direction of quintuple layer of
3D TI. $\Psi_{{\bf i}}=(a_{\bf i},b_{\bf i},c_{\bf i},d_{\bf i})^T$,
and $a_{\bf i},b_{\bf i},c_{\bf i},d_{\bf i}$ represents the four
annihilation operators of electron on the site ${\bf i}$ with the
state indices
$|P1^+_z,\uparrow\rangle,|P2^-_z,\uparrow\rangle,|P1^+_z,\downarrow\rangle,|P2^-_z,\downarrow\rangle$.
In Eq.(\ref{Ham}), $H_{{\bf ii}}$ and $H_{{\bf i},\vec{\delta} x}$
are $4\times4$ Hamiltonian that are given by $H_{{\bf
ii}}=(C+2D_1/a^2+4D_2/a^2)I_{4\times4}+(M-2B_1/a^2-4B_2/a^2)\Gamma_0$,
$H_{{\bf i},\vec{\delta}
x}=-D_2/a^2I_{4\times4}+B_2/a^2\Gamma_0-iA_2/(2a)I^a_{4\times4},
H_{{\bf i},\vec{\delta}
y}=-D_2/a^2I_{4\times4}+B_2/a^2\Gamma_0-iA_2/(2a)\Gamma_{1}, H_{{\bf
i},\vec{\delta}
z}=-D_1/a^2I_{4\times4}+B_1/a^2\Gamma_0-iA_1/(2a)\Gamma_{2}$, where
$\Gamma_{0,1,2}\equiv (I_{2\times2}\otimes s_z,s_y\otimes
I^a_{2\times2},s_z\otimes I^a_{2\times2})$ and $I^a$ is
anti-diagonal identity matrix. Static Anderson type disorder is
added to the on-site energy with a uniform distribution in the
interval $[-W/2,W/2]$ where W characterizes the strength of the
disorder. Here $a$ is the lattice constant, and $A_1, A_2, B_1, B_2,
C, D_1, D_2$ and $M$ are system's parameters taken from
Ref.\onlinecite{Zhang}.

\begin{figure}
\includegraphics[width=8.5cm,height=6.5cm]{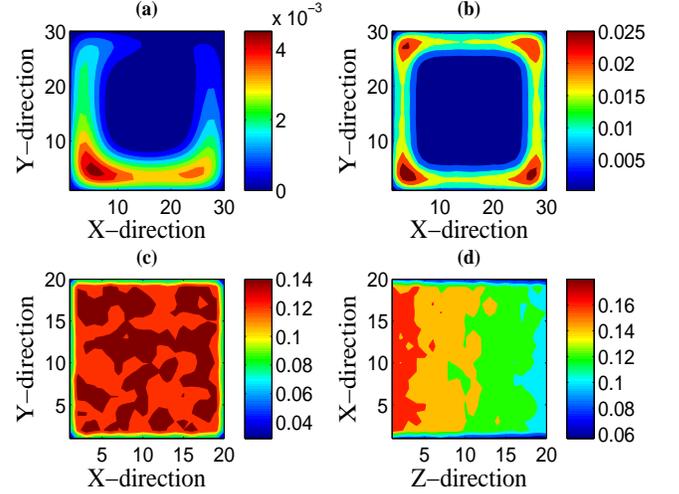} \caption{
(Color online)Averaged DOS in the middle slices of the
central simulation box with injected wave from left lead with
different disorder strength.\label{fig2}}
\end{figure}

By using Green's functions, the charge conductance from the
terminal-$\beta$ to the terminal-$\alpha$ can be calculated by using
Landauer-B\"{u}ttiker formula $G_{\alpha\beta}(E)=
(e^2/h)T_{\alpha\beta}$, and
$T_{\alpha\beta}=\mathrm{Tr}[\Gamma_{\alpha}G^r\Gamma_{\beta}G^a]$
is the transmission coefficient. The linewidth function
$\Gamma_{\alpha}(E)=i[\Sigma^r_{\alpha}- \Sigma^a_{\alpha}]$ and the
Green's functions $G^{r/a}(E)$ can be calculated from
$G^r=[G^a]^{\dagger}=[EI-H_C-\sum_{\alpha}\Sigma^r_{\alpha}]^{-1}$,
where $H_C$ is Hamiltonian matrix of the central scattering region
and $I$ is the unit matrix with the same dimension as that of $H_C$,
$\Sigma^r_{\alpha}$ are self energy of external leads and can be
calculated numerically\cite{self}. The conductance fluctuation is
defined as $\text{rms}(G)\equiv \sqrt{\left\langle
G^{2}\right\rangle -\left\langle G\right\rangle^{2}}$, where
$\left\langle {\cdots}\right\rangle $ denotes averaging over an
ensemble of samples with different disorder configurations of the
same strength $W$. In the following the average conductance and its
fluctuation are measured in unit of $e^2/h$.

In the numerical calculations, we choose the realistic material
parameters for Bi$_2$Se$_3$\cite{Zhang} and perform calculations on
a $L\times L\times L$ cubic sample with $L=20$ and the lattice constant
$a=5$ {\AA}. Since each site has four orbitals, the dimension of the
Hamiltonian or Green's function becomes $32000$ for this system
making the simulation very computational demanding. We first study
the two terminal device by considering different transport
directions in the presence of Anderson type disorder. As expected,
our numerical simulation shows that transport properties along x
direction is the same as that of y direction. So for the two-terminal
structure there are four conductances: $G_{zz}$, $G_{xx}$, $G_{xy}$, and
$G_{zx}$ that are plotted in Fig.1 along with their fluctuations
against the disorder strength. Here the Fermi energy is inside the
bulk gap so that only topological surface states are conducting channels.
In general, our results show that the topological
surface states are gradually destroyed in the presence of small
disorders. As the strength of disorder increases, the system enters
the diffusive regime and the conductance becomes quantized for a
wide range of disorder strength. Importantly, this quantization of conductance
is accompanied by the quantized conductance fluctuation. This
conductance fluctuation is universal since it is independent of
parameters such as Fermi energy and disorder
strength as shown in Fig.3. From Fig.1, we see that the quantized conductance takes
fractional value with $\langle G_{zz}\rangle=1$, $\langle G_{xx}\rangle=\langle G_{xy}\rangle=4/3$, and
$\langle G_{zx}\rangle =6/5$. Two points worth mentioning from Fig.1(c) and (d): (1). Two set of curves
for $\langle G\rangle$ and $\text{rms}(G)$ versus disorder strength along x direction and y direction are exactly the same which is expected. (2). For $G_{xx}$ and $G_{xy}$, however, only their average conductance and its fluctuation in the diffusive regime are the same. Its origin can be understood using the theory of mode mixing discussed below.
To make sure that the quantization plateau lies in the diffusive regime, we have calculated the localization length $\xi$ in the quantization plateau region for $W=[5,10]$. Our results give $\xi/L =6 \sim 8$ in this range of $W$
indicating that the system is indeed in the diffusive regime.

\begin{figure}
\includegraphics[width=8.5cm,height=6.5cm]{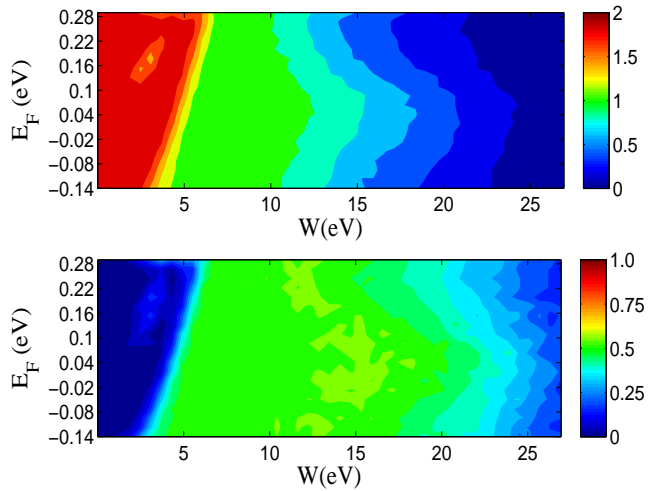} \caption{
(Color online) Phase diagram of conductance (panel (a)) and its fluctuation (panel (b))
in the plane of $(W,E_F)$. 1000 samples are collected in each point on the phase diagram
calculation.\label{fig3}}
\end{figure}

The quantization of conductance in the diffusive regime can be
described using the following phenomenological theory of mode mixing
in the central region.\cite{Levitov} For instance, for the transport
along z direction, two incoming topological surface states are
completely mixed in the scattering region in the diffusive regime
with equal probability of going forward $p_f$ and backward $p_b$,
i.e., $p_f=p_b=1/2$. Hence the conductance is $\langle G_{zz}\rangle=2 p_f=1$.
Alternatively, we can use the number of effective transmission channels
$\nu_t$ and reflection channels $\nu_r$ with $\nu_t/\nu_r=p_f/p_b$.
For $G_{zz}$, we have $\nu_r=\nu_f=1$ so that $p_f=\nu_t/(\nu_t+\nu_r)=1/2$.
For $G_{xx}$, only partial mixing is achieved with $\nu_r=1$ and
$\nu_t=2$, or $p_f=2/3$. As a result, $\langle G_{xx}\rangle=2 p_f = 4/3$. Since x and y directions are equivalent, an incoming electron from x direction traversing in the scattering region find the same available $\nu_t$ along y direction as that of x direction. Hence $\langle G_{xx}\rangle=\langle G_{xy}\rangle$. To support this argument, we have calculated two-terminal conductance of a four-terminal device with four terminal on x-y plane (see the inset of Fig.4(b)). For an electron from L-lead, we should have $\langle G_{RL}\rangle=\langle G_{UL}\rangle=\langle G_{DL}\rangle$ if x and y directions are equivalent. This is indeed what we found numerically. In addition, the conductance fluctuations are also the same. For $G_{zx}$, $\nu_r=1$ and $\nu_t=3/2$ due to the partial mixing, or $\langle G_{zx}\rangle=2 p_f = 6/5$. In terms of the number of effective
channels, the quantized conductance can be given by the following
expression,
\begin{eqnarray}
\langle G\rangle=2\frac{\nu_t\nu_r}{\nu_t+\nu_r}.\label{mean}
\end{eqnarray}
This ansatz gives the average conductance that agrees with our
numerical result for the quantized conductance of
two-terminal structures. The universal conductance fluctuations(UCF)
can also be expressed in terms of the number of effective transmission
and reflection channels,\cite{foot}
\begin{eqnarray}
\mathrm{rms}(G)=4\sqrt{\frac{\nu_t^2\nu_r^2}{(\nu_t+\nu_r)^2[4(\nu_t+\nu_r)^2-2]}}.\label{var}
\end{eqnarray}
With $(\nu_r,\nu_f)=(1,1)$, $(1,2)$, and $(1,1.5)$ for $G_{zz}$,
$G_{xx}$, and $G_{zx}$, respectively, we find from Eq.(\ref{var})
that $\mathrm{rms}(G_{zz})=2/\sqrt{14}=0.535$,
$\mathrm{rms}(G_{xx})=8/3/\sqrt{34}=0.457$, and
$\mathrm{rms}(G_{zx})=12/5/\sqrt{23}=0.500$ that are very close to
our numerical results (see Fig.1).

\begin{figure}
\includegraphics[width=8.5cm,height=6.5cm]{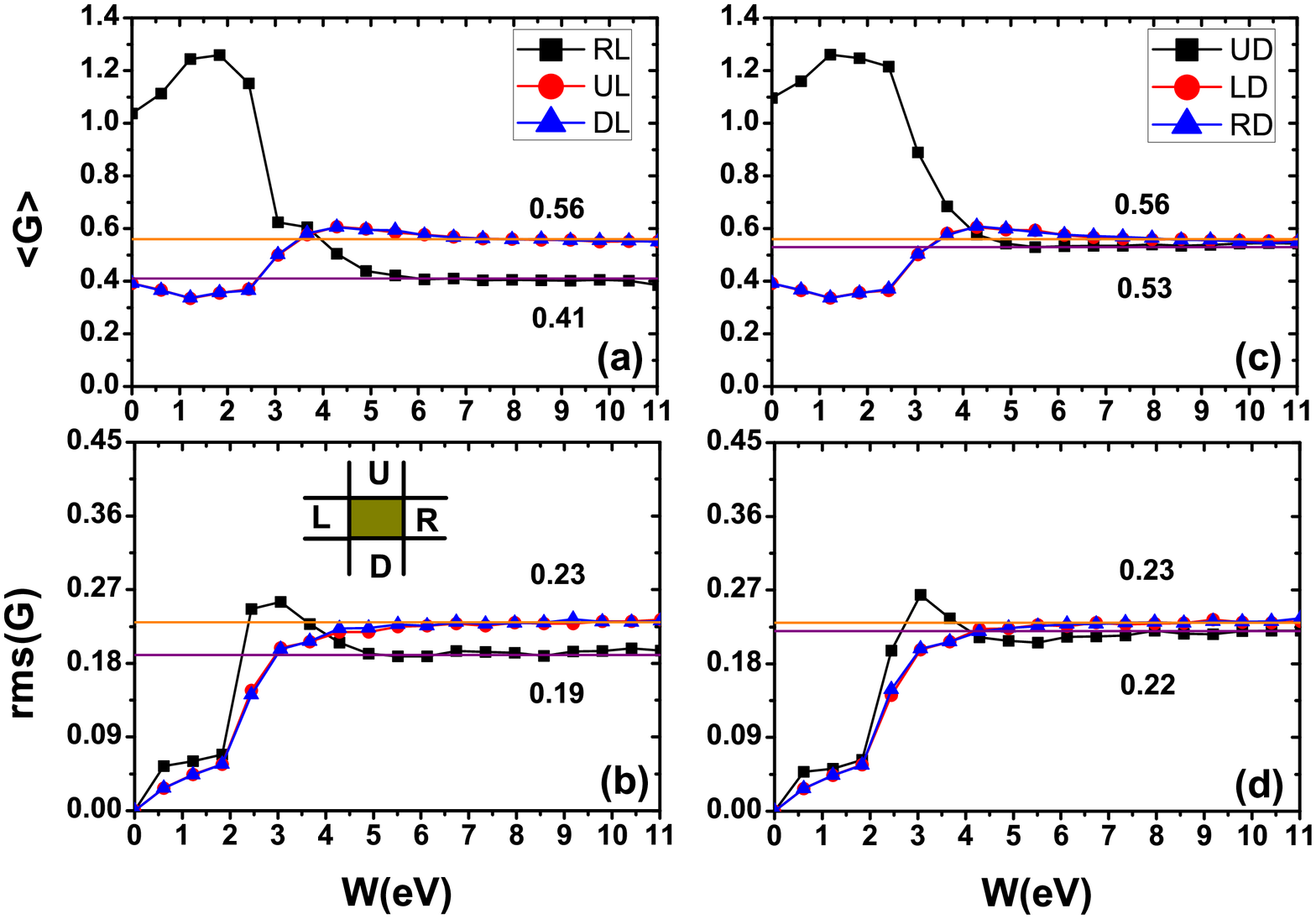} \caption{
(Color online) Conductance and its fluctuation vs disorder strength
for different transport direction in four terminal
cross-setup.\label{fig4}}
\end{figure}

We now examine the evolution of wave function as disorder strength
is varied in the diffusive regime when the propagation is along z
direction. In our numerical simulation we set $E_F=0.1348eV$ in Fig.2(a),(b)
and $E_F=0.0178eV$ in Fig.2(c),(d).
In the clean sample there are two pairs of topological surface states localized
on the surface and separated in space so that the backscattering is
forbidden. In Fig.2(a), we have shown the DOS of one of the topological surface states propagating along positive z direction. As the disorder is increased, these localized surface
states become extended on the surface and the topological surface
states are destroyed. However, current carrying states are still
surface states and there is no DOS in the bulk (see Fig.2(b)). This behavior of
bulk insulator persists when the average conductance becomes
quantized in the diffusive regime, e.g., when $W=6eV$ the system is a bulk insulator but
the surface states is not topological protected. As disorder strength increases
further the surface states gradually expand toward the center of the
system and eventually become bulk states. At $W=8eV$, the bulk gap is closed
due to the disorders. In order to investigate the nature of mode mixing,
we calculate the DOS for electron coming from the left lead along z direction
at $W = 8 eV$. Fig.2(c) shows the average DOS in $x-y$ plane where $10000$ configurations
are collected. From Fig. \ref{fig2}, we see that two modes are fully mixed in $x-y$
plane due to the disorder scattering. In the $x-z$ plane (Fig.2(d)), the DOS is
mainly distributed around the left region due to the suppression of
transmission ($\langle G\rangle\simeq1$).

To demonstrate the universal feature of the quantized conductance and its
fluctuation, we have calculated the phase diagram where $G_{zz}$ and $\text{rms}(G_{zz})$ are plotted in
(W,$E_F$) plane (see Fig.3). In the calculation, the periodic condition in $x$ direction is
employed. The light green region shows the quantized conductance. For
Fermi energy inside the energy gap in the clean sample, we see that
the system goes quickly from topological insulator to the region of quantized
conductance. As disorder strength increases, it slowly enters the localized
regime. This phase diagram shows that the conductance and its fluctuation
are independent of Fermi energy and disorder strength in a wide range.

Now we study the transmission coefficient $T_{\alpha \beta}$ of a four-terminal device shown schematically
in the inset of Fig.4(b) where the leads L and R are along z direction and leads
U and D are along x direction. In the calculation we have fixed $E_F=0.0178eV$ and over $5000$ samples are collected for each average. Fig.4 shows that in the diffusive regime, the conductance and its fluctuation are again quantized. When there are four terminals, however, the quantized conductance is no longer a fractional value. It is much smaller than that of the two-terminal case. This is not unexpected since there are two more terminals that electron can exit. To our surprise, the conductance fluctuation is also much smaller than that of the two-terminal case, nearly halved.

To provide further evidence of universal conductance fluctuation, we have performed calculation for another 3D TI in $Sb_2Te_3$\cite{liu} on a $20\times 20\times 20$ cubic sample. Our numerical results give the same values of quantized conductance and fluctuation. So far, we have studied the influence of bulk disorder on the transport of 3D TI. We have also calculated the averaged conductance and its fluctuation in the presence of "surface" disorder. Specifically, we have calculated $G_{zz}$ when disorders are present only on the first layer or first three layers of surface of 3D TI in $Bi_2Se_3$. The quantized conductance and its fluctuation again show the same quantized plateau as the case of the bulk disorder with the same quantized conductance and quantized fluctuation. Note that the conductance measurement with surface disorder has been carried out experimentally on $Bi_2 Se_3$.\cite{chen,he,wang} Since the quantization of conductance and fluctuation exist in a large window of disorder strength (corresponding to doping concentration in experiment), we believe that our results can be checked experimentally.

To summarize, we have carried out extensive numerical simulation to calculate the transport properties of disordered 3D TI. Our results show that in the diffusive regime, the two-probe conductance is quantized with fractional value $1,4/3,6/5$. The corresponding conductance fluctuation is also quantized. An effective mode mixing theory is proposed that gives quantized conductance and fluctuation in excellent agreement with our numerical results. The numerical results from different parameters including disorder strength and Fermi energy (from phase diagram of conductance and its fluctuation), types of disorders (bulk and surface disorder), and types of 3D TI ($Bi_2Se_3$ and $Sb_2Te_3$) suggest that the quantized conductance and quantized fluctuation is a universal property of 3D TI. We have also studied effect of disorder on topological surface states. We found that as disorder strength increases the conducting channels changes from topological surface states to non-topological surface states and finally to bulk states in the diffusive regime.

{\bf Acknowledgments} This work was financially supported by Research Grant Council
(HKU 705409P) and University Grant Council (Contract No. AoE/P-04/08) of the Government of HKSAR. This
research is conducted using the HKU Computer Centre research
computing facilities that are supported in part by the Hong Kong UGC
Special Equipment Grant (SEG HKU09).

$^*$ Electronic address: jianwang@hku.hk


\begin{thebibliography}{99}
\bibitem{foot}
Note that this expression is different from Eq.(4) in Ref.\onlinecite{Levitov}.
\end{thebibliography}
\end{document}